\documentclass[prl, two column,showkeys,showpacs]{revtex4-1}
\date{today}
\usepackage{natbib,hyperref}
\usepackage{graphicx}
\usepackage{epstopdf}
\usepackage{amsmath}
\usepackage{changes}
\usepackage{listings}
\usepackage{color}
\definecolor{codegreen}{rgb}{0,0.6,0}
\definecolor{codegray}{rgb}{0.5,0.5,0.5}
\definecolor{codepurple}{rgb}{0.58,0,0.82}
\definecolor{backcolour}{rgb}{0.95,0.95,0.92}
\lstdefinestyle{mystyle}{
    backgroundcolor=\color{backcolour},
    commentstyle=\color{codegreen},
    keywordstyle=\color{magenta},
    numberstyle=\tiny\color{codegray},
    stringstyle=\color{codepurple},
    basicstyle=\footnotesize,
    breakatwhitespace=false,
    breaklines=true,
    captionpos=b,
    keepspaces=true,
    numbers=left,
    numbersep=5pt,
    showspaces=false,
    showstringspaces=false,
    showtabs=false,
    tabsize=2}
\newcommand{\ket}[1]{\left|#1\right>}
\newcommand{\bra}[1]{\left< #1 \right|}

\newcommand{\beq}{\begin{equation}}
\newcommand{\eeq}{\end{equation}}
\newcommand{\beqa}{\begin{eqnarray}}
\newcommand{\eeqa}{\end{eqnarray}}

\begin{document}
\title{Spin decoherence in a two-qubit CPHASE gate: the critical role of tunneling noise}
%\title{Spin deoherence in a two-qubit logic gate: critical role of tunneling fluctuation}

%\title{Decoherence of a Two-Qubit Gate in Silicon Quantum Dots}
\author{Peihao Huang$^{1,2,3}$}
%\email{peihao.huang@nist.gov}
\email{phhuang.cmp@gmail.com}
\author{Neil M. Zimmerman$^{2}$}
\author{Garnett W. Bryant$^{1,2}$}
\affiliation{$^1$Joint Quantum Institute, National Institute of Standards and Technology and University of Maryland, Gaithersburg, Maryland, 20899}
\affiliation{$^2$Quantum Measurement Division, National Institute of Standards and Technology, Gaithersburg, Maryland, 20899}
\affiliation{$^3$Shenzhen Institute for Quantum Science and Engineering, and Department of Physics, Southern University of Science and Technology, Shenzhen 518055, China}
%\affiliation{$^3$Department of Physics, University at Buffalo, SUNY, Buffalo, New York 14260}

\date{\today}

%Neil Zimmerman, Dimitrie Culcer, Xuedong Hu$^3$,
%Michael Gullans, Jake Taylor, Charles Tahan

\begin{abstract}
Rapid progress in semiconductor spin qubits has enabled experimental demonstrations of a two-qubit logic gate.
Understanding spin decoherence in a two-qubit logic gate is necessary for optimal qubit operation.
We study spin decoherence due to $1/f$ charge noise for two electrons in a double quantum dot used for a two-qubit controlled-phase gate. In contrast to the usual belief, spin decoherence can be dominated by the tunneling noise from $1/f$ charge noise instead of the detuning noise. Tunneling noise can dominate because the effect of tunneling noise on the spin qubit is first order in the charge admixture; while the effect of the detuning noise is only second order. The different orders of contributions result in different detuning dependence of the decoherence, which provides a way to identify the noise source. We find that decoherence in a recent two-qubit experiment was dominated by the tunneling noise from $1/f$ charge noise.
The results illustrate the importance of considering tunneling noise to design optimal operation of spin qubits.
%Understanding decoherence in a two-qubit gate is essential for the better control of spin qubit.

\end{abstract}

%\keywords{semiconductor; quantum dot (QD); quantum information; spin qubit; two-qubit gate; CPHASE gate; S-T$_0$ qubit; charge noise; $1/f$ noise; dephasing; decoherence}
%\pacs{}

\maketitle

\section{introduction}

An electron spin confined in a semiconductor quantum dot (QD) is a promising candidate quantum bit (qubit) for quantum information processing because of its potential scalability and miniaturization \cite{loss1998, kane1998}.
Tremendous progress has been made during the last decade \cite{petta2005, hanson_spins_2007,maune_coherent_2012,veldhorst_addressable_2014,kim_microwave-driven_2015, eng_isotopically_2015, cao_tunable_2016, kawakami_gate_2016, yoneda2018}.
%, including single qubit operation of a single spin in GaAs \cite{koppens_driven_2006, nowack_coherent_2007} or silicon \cite{maune_coherent_2012, veldhorst_addressable_2014, kawakami_electrical_2014, kawakami_gate_2016, yoneda2018}, multiple-axis operation \cite{petta2005, foletti_universal_2009, van_weperen_charge-state_2011, medford_scaling_2012, medford_quantum-dot-based_2013, medford_self-consistent_2013, shi_fast_2014, kim_microwave-driven_2015, eng_isotopically_2015, cao_tunable_2016} and two-qubit entanglement \cite{shulman_demonstration_2012,  nichol_high_fidelity_2017} of encoded spin qubits (such as singlet-triplet ($S-T_0$) qubits) in GaAs.
The spin qubit in a silicon QD has attracted wide interest because of its long coherence time and compatibility with Si electronics nanofabrication \cite{morton_embracing_2011, zwanenburg_silicon_2013}. With advances in fabricating QDs using accumulation mode \cite{zajac_reconfigurable_2015}, several groups have demonstrated two-qubit gates in silicon based on the exchange interaction \cite{veldhorst2015, zajac_resonantly_2018, watson_programmable_2018}. The recent achievement of strong coupling between spin qubits and microwave photons will also enable long distance gate operations for spin qubits \cite{viennot_coherent_2015, mi_coherent_2018, samkharadze_strong_2018}.

%(The CPHASE and CNOT gate demonstrated recently between two electrons in the neighboring QDs are based on the exchange interaction.)
Two-qubit gate operation is an essential but challenging task for the building of a quantum computer. For spin qubits, there are many proposals for gates in the literature, including two-qubit gates based on exchange interaction, strong coupling of spin and photon, superexchange coupling, and capacitive coupling \cite{veldhorst2015, zajac_resonantly_2018, watson_programmable_2018, mi_coherent_2018, samkharadze_strong_2018, baart_coherent_2017, nichol_high_fidelity_2017}.
In recent experiments in silicon, the two-qubit gates, including controlled-phase (CPHASE) gate and controlled-not (CNOT) gate, are mediated by the exchange interaction between electrons in the nearby QDs \cite{veldhorst2015, zajac_resonantly_2018, watson_programmable_2018}. In this study, we focus on a CPHASE gate mediated by the exchange interaction, which was studied experimentally \cite{veldhorst2015}. Most of the results will also apply to other two-qubit gates based on the exchange interaction.

%There are different types of two-qubit gates studied in the literature for spin qubits. The controlled-phase (CPHASE) and controlled-not (CNOT) gate have been demonstrated between electrons in the neighboring QDs based on exchange interaction. There are also proposals for two-qubit gates based on the spin-photon coupling and moving electrons.
%The two-qubit gate we are interested in in this paper is a CPHASE gate based on the exchange interaction. Most of the results in this paper will also apply to the other two-qubit gates based on the exchange interaction, such as a direct CNOT gate -cite{xxx}.

%, tyryshkin_electron_2012
Environmental noise can destroy the quantum coherence necessary for quantum computation.
For a spin qubit in silicon, nuclear noise can be suppressed with isotopic enrichment \cite{tyryshkin_electron_2003}.
However, low frequency $1/f$ charge noise remains ubiquitous in solid state devices. Consequently, spin dephasing is typically dominated by charge noise \cite{hu2006, culcer_dephasing_2013}.
For QD devices, charge noise can be measured by monitoring the charge offset drift in a single electron transistor, where noise causes fluctuation of the electron chemical potential $\delta \mu$ in a single QD and shifts the Coulomb blockade spectrum \cite{zimmerman_why_2008}. The power spectral density
$S_{1/f}(\omega)=\int_{-\infty}^{\infty}\langle \delta \mu (0) \delta \mu(\tau) \rangle \cos(\omega\tau)d\tau$
of energy fluctuation due to charge noise is typically
\beq
S_{1/f}(\omega) = A/\omega,
\eeq
where $A$ is the amplitude of charge noise \cite{dutta_lowfrequency_1981, weissman_frac1f_1988}. Although $S_{1/f}(\omega)$ usually exhibits $1/\omega$ dependence, the exponent of $\omega$ can be different for different devices. The energy fluctuations $\sqrt{A}$ ranges from 0.1 $\mu$eV to 10 $\mu$eV depending on material and experimental details \cite{dutta_lowfrequency_1981, weissman_frac1f_1988, zimmerman_charge_2014, freeman_comparison_2016}.
%In this study, we focus on the effect of charge noise on spin decoherence and the number of gate operations in a two-qubit CPHASE gate of two electron spin qubits in a double quantum dot (DQD).

For two electrons in a double quantum dot (DQD), where the spins can be used to define a single singlet-triplet (S-T0) qubit or two single-spin qubits coupled by exchange interaction, the $1/f$ charge noise can significantly affect the spin coherence and the number of gate operations.
Usually, the detuning fluctuation from charge noise is assumed to be the dominant source of decoherence \cite{stopa_magnetic_2008, culcer_dephasing_2009, nielsen_implications_2010, yang_low-noise_2011, kalra_robust_2014}. When that is the case, symmetric operation can increase the number of qubit operations as demonstrated in recent experiments \cite{bertrand_quantum_2015, reed_reduced_2016, martins_noise_2016}. Furthermore, many theoretical papers on the optimal operation of spin qubit are based on reducing the sensitivity to detuning fluctuation from charge noise \cite{russ_asymmetric_2015, shim_charge-noise-insensitive_2016, zhang_randomized_2017, yang_suppression_2017, friesen_decoherence-free_2017}.
However, the assumption that spin decoherence is dominantly due to detuning fluctuation may not always be satisfied, especially for spin qubits in small silicon QDs formed in accumulation mode. In this work, we show that tunneling fluctuation can play an important role for two electrons in a DQD used for a two-qubit CPHASE gate. The dominance of tunneling noise can significant modify the optimal operation of a spin qubit.

%Not much attention has been paid to tunneling fluctuation from charge noise and its role in small silicon QDs using accumulation mode.

%\textit{System Hamiltonian---}
{
%In the regime, where the ground state charge configuration is $(1,1)$,
We consider two electrons in a gate-defined DQD for a two-qubit CPHASE gate mediated by the exchange interaction.
First, we introduce the Hamiltonian that describes two electrons in a DQD.
The system Hamiltonian is $H=H_Z+H_C$, where $H_Z$ is the Zeeman term, and $H_C$ is the Hamiltonian for the charge degree of freedom
\cite{veldhorst2015}.
In the case of a silicon DQD, we assume that the electron Zeeman splitting and thermal energy are well below the valley splitting, so that we consider only the lowest valley state \cite{takashina_valley_2006, goswami_controllable_2007, yang2013,  hao_electron_2014}.
%The strong interface field lifts the valley degeneracy, and splits the lowest two valley states by a few hundred $\mu$eV \cite{takashina_valley_2006, goswami_controllable_2007, yang2013,  hao_electron_2014}.
%Assuming that the electron Zeeman splitting and thermal energy are well below the valley splitting, we consider only the lowest valley state \cite{veldhorst2015}.
%The charge configuration of the DQD can be labeled as $(N_l,N_r)$, where $N_l$ ($N_r$) is the number of electrons in the left (right) QD. The basis states can be denoted as $\ket{(N_l,N_r)X}$, where $X$ is the two-electron spin state ($S$ for spin singlet and $T_+$, $T_0$, or $T_-$ for spin triplet).
The basis states can be denoted as $\ket{(N_l,N_r)X}$, where $N_l$ ($N_r$) is the number of electrons in the left (right) QD, $X$ is the two-electron spin state ($S$ for the spin singlet and $T_+$, $T_0$, or $T_-$ for the spin triplet).
In the (1,1) charge regime, the lowest four states $\ket{(1,1)X}$ are degenerate in the absence of magnetic field, and the energies of the double occupation states $\ket{(2,0)S}$ and $\ket{(0,2)S}$ are higher because of the strong electron-electron repulsion.
$\epsilon$ is the detuning defined relative to the symmetric operation point. The energy of the double occupation states can be raised or lowered by the detuning. The tunneling $t$ between the two dots can couple $\ket{(1,1)S}$ to the double occupation states. Figure 1(a) shows the energy diagram of the relevant singlet states, and the coupling between the
states.
%Figure \ref{Fig_scheme} (a) shows the energy diagram of the relevant singlet states with detuning $\epsilon_0$ defined relative to the symmetric point of the DQD.
%The tunneling $t_{0}$ couples the single occupation singlet state $\ket{(1,1)S}$ to the double occupation state $\ket{(2,0)S}$ (or $\ket{(0,2)S}$), and the energy cost for the tunneling is $U-\epsilon_{0}$ (or $U+\epsilon_{0}$), where $U$ is the energy due to Coulomb interaction.
For a two-qubit logic gate in Ref. \cite{veldhorst2015}, which is the experiment most relevant to our theory, the DQD is operated in the asymmetric regime (large $\epsilon$) of the $(1,1)$ charge regime.
We need to consider only the lower energy double occupation state, shown as $\ket{(2,0)S}$ in Figure 1(a) with the left dot at lower energy. Then, the system Hamiltonian is
\beqa
H_Z &=& \overline{E}_Z\sum_m m \ket{(1,1)T_m}\bra{(1,1)T_m} \nonumber \\
&+& \frac{\delta E_Z}{2} \left(\ket{(1,1)T_0}\bra{(1,1)S} + h.c. \right),\label{H_Z}\\
H_{C} &=& ({U}-\epsilon) (\ket{(2,0)S}\bra{(2,0)S})\nonumber \\
&+& \sqrt{2}t_{} \left(\ket{(1,1)S}\bra{(2,0)S} + h.c. \right) , \label{H_C}
%H_Z &=& \overline{E}_Z\sum_m m \ket{T_m}\bra{T_m} + \frac{\delta E_Z}{2} \left(\ket{T_0}\bra{S} + h.c. \right),\label{H_Z}\\
%H_{C}&=& \sqrt{2}t_{} \left(\ket{S}\bra{S_{D}} + h.c. \right) + ({U}-\epsilon) (\ket{S_D}\bra{S_D}), \label{H_C}
%\lefteqn{H_{C} = \sqrt{2}t_{} \left(\ket{(1,1)S}\bra{(2,0)S} + h.c. \right) } \\
%&& + ({U}-\epsilon) (\ket{(2,0)S}\bra{(2,0)S}), \label{H_C}
\eeqa
where $\delta E_Z=(E_{Z,l}-E_{Z,r})$ (or $\overline{E}_Z=(E_{Z,l}+E_{Z,r})/2$) is the difference (or average) in the Zeeman splitting of the left (l) and right (r) dots, the index m=-1,0, or 1, and $U$ is the energy due to the electron-electron repulsion.

%where $t=t_0 + \hat{n}_t$ and $\epsilon=\epsilon_0 + \hat{n}_\epsilon$ are the tunneling and detuning in the presence of noise.
%In $H_Z$,
%${U}$ is the onsite Coulomb potential (i.e. charging energy of the left QD), $t_0$ is the hopping amplitude that couples $\ket{(1,1)S}$ and $\ket{(2,0)S}$, and $\epsilon$ is the detuning of the DQD.
%For the noise Hamiltonian $H_1$, we will focus on low frequency $1/f$ charge noise, which can induce strong dephasing in a spin qubit or superconducting qubit. \cite{astafiev_quantum_2004, makhlin_dephasing_2004, simmonds_decoherence_2004, martinis_decoherence_2005, ithier_decoherence_2005, hu2006}.
%Although, the origin of charge noise can be different in different systems, it is generally believed that charge noise is from fluctuating two-level systems in the material. \cite{anderson1972, phillips1972, phillips1987, simmonds_decoherence_2004, martinis_decoherence_2005,zimmerman_why_2008}
%%For example, in semiconductor QDs, the fluctuation can be the result of filling (emptying) electrons into (from) a trap at an interface, dipole defects in the material or charged ions at the surface. \cite{zimmerman_why_2008}

\begin{figure}[t]
\centering
\includegraphics[scale=0.85]{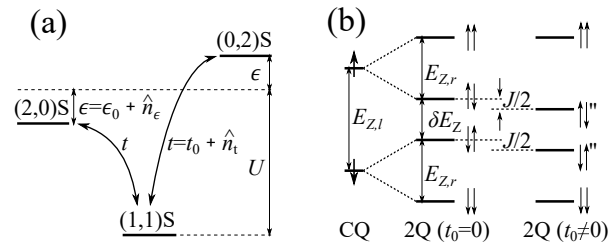}
\caption{Schematic energy diagrams. (a): Energy diagram of singlet states for two electrons in a DQD. Detuning $\epsilon=\epsilon_0+\hat{n}_\epsilon$ and tunneling $t=t_{0} + \hat{n}_t$ are indicated, in which $\hat{n}_\epsilon$ ($\hat{n}_t$) is detuning (tunneling) fluctuation from charge noise. (b): The energy diagram of control-qubit (CQ) states, two-qubit (2Q) states without exchange interaction (i.e. $t_{0}=0$), and two-qubit eigenstates with exchange
interaction $J$.
%The control-qubit state selects the subspace of two-qubit system, and $J$  determines the difference of Z-rotation speeds of target qubit for different control-qubit states.
%(c). Effect of $1/f$ charge noise on spin decoherence in a two-qubit logic gate.
}\label{Fig_scheme}
\end{figure}

%Schematic of charge stability diagram. Charge configuration $(N_L,N_R)$ is indicated in each regime, where $N_L$ ($N_R$) is the number of electrons in the left (right) QD. Detuning $\epsilon$ (related to asymmetry) is defined with respect to the line of symmetric operation of a DQD as indicated; (b)
%(c) Schematic diagram of chemical potentials of two electrons in a DQD. Charge noise causes fluctuations of both the detuning $\epsilon$ and the tunneling  $t_0$;

Here, we include noise due to charge fluctuation.
The electrical field due to charge noise causes a fluctuation of the electrical potential of each QD, and thus leads to the detuning noise $\hat{n}_{\epsilon}$ of a DQD, $\epsilon = \epsilon_0 + \hat{n}_\epsilon$.  $\hat{n}_\epsilon$ is usually considered as the decoherence source of a spin qubit in a DQD \cite{stopa_magnetic_2008, culcer_dephasing_2009, nielsen_implications_2010, yang_low-noise_2011, kalra_robust_2014, shim_charge-noise-insensitive_2016, zhang_randomized_2017, friesen_decoherence-free_2017}.
%%-cite{S-T0decoherence}
Charge noise also causes fluctuation of the tunneling barrier and results in tunneling noise $\hat{n}_t$, $t = t_0 + \hat{n}_t$.
%(see Figure \ref{Fig_scheme} (c))
Here, $\hat{n}_\epsilon$ and $\hat{n}_t$ vary randomly in time. The fluctuation of both $\epsilon$ and $t$ provides a mechanism for spin dephasing as show below.
%not operators if classical noise is considered, and they
%Therefore, the noise Hamiltonian $H_1$ {in the basis $B_{2Q}$} is given by
%Thus, both detuning noise and tunneling noise from $1/f$ charge noise can cause spin decoherence. %(see Figure \ref{Fig_scheme} (c)).

The Hamiltonian without noise describes the physics of a two-qubit controlled logic gate using the two electrons in the DQD based on the exchange interaction as we now describe. The electron spin in the left (right) QD act as a control-qubit (target-qubit).
The qualitative energy diagram of the two-qubit states is shown in Figure \ref{Fig_scheme} (b).
%Fig. \ref{Fig_scheme} (d) shows qualitatively the energy diagram of spin states in the absence of noise.
%Before solving the problem, we discuss qualitatively the spin states in the system. Fig. \ref{Fig_scheme} (d) shows the energy diagram of spin states $\ket{(1,1)X}$ in the presence of an applied magnetic field.
%In the (1,1) charge regime,
If there is no tunneling, $t_0=0$, then, $H_Z$ alone governs the spin degree of freedom.
For a finite difference in the Zeeman splitting $\delta E_Z$ at two dots, the four spin eigenstates are $\ket{\uparrow\uparrow}=\ket{(1,1)T_+}$, $\ket{\uparrow\downarrow}=\ket{(1,1)T_0}+\ket{(1,1)S}$, $\ket{\downarrow\uparrow}=\ket{(1,1)T_0}-\ket{(1,1)S}$, and $\ket{\downarrow\downarrow}=\ket{(1,1)T_-}$.
The energy splitting of the target-qubit does not depend on the control-qubit's state and there is no two-qubit gate.
However, if $t_0 \ne 0$, the tunnel coupling results in an effective exchange interaction $J$, which lowers the energies of states $\ket{\uparrow\downarrow}$ and $\ket{\downarrow\uparrow}$ \cite{meunier_efficient_2011, veldhorst2015}.
As a consequence, the energy splitting of the target-qubit depends on the state of the control-qubit, allowing the two spins to operate as a controlled logic gate. Then, a Ramsey-like pulse sequence will result in a two-qubit CPHASE gate between the two electron spin qubits \cite{meunier_efficient_2011, veldhorst2015}.

The tunneling fluctuation and the detuning fluctuation from the $1/f$ charge noise will cause the spin decoherence during the CPHASE gate operation. In the following, we focus on the decoherence and the number of operations, and show the importance of tunneling fluctuation in a recent two-qubit gate experiment.

\section{Results}
%\section{Effective Hamiltonian}
In this section, we first summarize the derivation of an effective Hamiltonian allows us to identify how the noises appears during gate operation in the (1,1) charge regime. More details are provided in the supplementary information S1. Then, we present the decoherence formula, the spectral density of the effective noise, and the spin decoherence results.

\textbf{Effective Hamiltonian---}
When $t_0 \ll U-\epsilon_0$, which is generally satisfied in experiments, an effective two-qubit Hamiltonian can be obtained for the system. The way to obtain the effective Hamiltonian is through an Schrieffer-Wolff transformation that decouples the higher energy states. Since we are considering only one of double occupation states, the procedure can be simplified. Consider the Hamiltonian without noise.
The Hamiltonian $H_C$ without noise can be diagonalized and the corresponding eigenstates can be denoted as $\ket{(1,1)S^\prime}$ and $\ket{(2,0)S^\prime}$. Here the prime denotes that the new state is close to the original state. The higher eigenstate $\ket{(2,0)S^\prime}$ is decoupled from the rest of the Hamiltonian up to the first order of $\frac{t_0}{U-\epsilon_0} \frac{\delta E_Z}{U-\epsilon_0}$ (Supplementary information S1), which is in general small. The lower eigenstate $\ket{(1,1)S^\prime}$ is approximately $\ket{(1,1)S}$ with certain admixture from $\ket{(2,0)S}$,
\beq
\ket{(1,1)S^\prime} \approx \ket{(1,1)S} + \frac{\theta}{2} \ket{(2,0)S},
\eeq
where the admixture factor $\theta = -2\sqrt{2}t_0/(U-\epsilon_0)$.
%Since we are considering only one of double occupation states, the procedure can be simplified. Let's consider the Hamiltonian without noise. Since $\ket{(2,0)S}$ is only coupled to $\ket{(1,1)S}$, once the Hamiltonian $H_C$ is diagonalized, the higher eigenstate is decoupled from the rest of the Hamiltonian up to the first order of $\frac{t_0}{U-\epsilon_0} \frac{\delta E_Z}{U-\epsilon_0}$, which is in general small.
%If we first consider the Hamiltonian without noise, and effective two-qubit Hamiltonian can be obtained by decouple the double occupation states from the rest of the two-qubit states. Since we are considering only $\ket{(2,0)S}$ of the double occupation states, which couples to state $\ket{(1,1)S}$ as described in $H_c$. Once the Hamiltonian $H_c$ is diagonalized, then, the higher eigenstate $\ket{(2,0)S^\prime}$ is decoupled from the rest of the Hamiltonian up to the 2nd order of ....
%The lower eigenstate $\ket{(1,1)S^\prime}$ is approximately $\ket{(1,1)S}$ with certain admixture from $\ket{(2,0)S}$: $\ket{(1,1)S^\prime} \approx \ket{(1,1)S} + \frac{\theta}{2} \ket{(2,0)S}$, where the admixture factor $\theta = -2\sqrt{2}t_0/(U-\epsilon_0)$.
The charge admixture lowers the energy of $\ket{(1,1)S^\prime}$, which results in an effective exchange interaction $J$.
%
%The diagonalization of $H_C$ leads to the charge admixture of $\ket{S}$ and $\ket{S_D}$,
%The Hamiltonian $H_C$ without noise is first diagonalized.
%The tunneling couples $\ket{(1,1)S}$ and $\ket{(2,0)S}$ and admixes the two states, where admixture is proportional to $t_{0}/(U-\epsilon_{0})$.
%, which is the basis for most two-qubit operation schemes of spin qubits \cite{loss1998, kane1998}.
Due to this charge admixture, the charge noise also couples to the qubit subspace.
%, an effective noise term appears in the qubit subspace
$\ket{(1,1)T_+}$ and $\ket{(1,1)T_-}$ are decoupled from $\ket{(1,1)T_0}$ and $\ket{(1,1)S^\prime}$, so, the effective Hamiltonian for the states ($\ket{(1,1)T_0}$, $\ket{(1,1)S^\prime}$) that are affected by noise is
\begin{eqnarray}
H^\prime&=&\left[\begin{array}{ccccc}
0 & \frac{\delta E_Z}{2} \\ %V_T and V_S is set to zero now
\frac{\delta E_Z}{2} & J+\hat{n}_{}^\prime \\
\end{array}\right],\\
\hat{n}_{}^\prime &=& \sqrt{2}{{\theta}}\hat{n}_{t} - ({{\theta}}^2/4)\hat{n}_{\epsilon},
\end{eqnarray}
%where
%where $\hat{n}_{}^\prime=\bra{S^\prime} \sqrt{2}\hat{n}_{t} \ket{S}\bra{S_D} + \hat{n}_{\epsilon} \ket{S_D}\bra{S_D} \ket{S^\prime}$ is
%\beq
%\hat{n}_{}^\prime = \sqrt{2}{{\theta}}\hat{n}_{t} - ({{\theta}}^2/4)\hat{n}_{\epsilon},
%\eeq
where $J=\bra{S^\prime} H_C \ket{S^\prime} \approx - \frac{2t_0^2}{U-\epsilon_0}$ is the exchange interaction, and the noise
$\hat{n}_{}^\prime $ is formally $\hat{n}_{}^\prime=(\partial{J}/\partial t_0) \hat{n}_t + (\partial{J}/\partial \epsilon_0) \hat{n}_\epsilon$.
{Both tunneling noise $\hat{n}_t$ and detuning noise $\hat{n}_\epsilon$ act on the spin qubit subspace but in different orders of the admixture factor $\theta$.
\textit{The effective noise due to the tunneling fluctuation is first order in the small charge admixture,  $\partial J/\partial t_{0} \propto \theta$; while the noise due to the detuning fluctuation is a second order effect, $\partial J/\partial \epsilon_0 \propto \theta^2$.}
%The effective noise on the spin qubit from the tunneling noise is a 1st order effect of admixture, while the effective noise from detuning noise is 2nd order.
Tunneling noise is a first order effect because tunneling noise $\sqrt{2}\hat{n}_{t} \ket{(1,1)S}\bra{(2,0)S}$ is coupled to $\ket{(1,1)S}$ in the qubit subspace even without charge admixture, while $-\hat{n}_{\epsilon} \ket{(2,0)S}\bra{(2,0)S}$ is decoupled from the qubit subspace and requires admixture to couple to $\ket{(1,1)S}$. Even if the tunneling noise is smaller than the detuning noise, which is generally true as shown below, the effect of tunneling noise may still be dominant.
}

%If DQD is detuned so that $\ket{(0,2)S}$ has much higher energy than $\ket{(2,0)S}$, then, admixture happens mainly between $\ket{(1,1)S}$ and $\ket{(2,0)S}$ and

%The same admixture also causes a spin qubit to be sensitive to charge noise.
%\textit{The sensitivity of exchange interaction to the tunneling fluctuation $ \partial J/\partial t_{0} \propto \theta$ is a first order effect in the small charge admixture, while the sensitivity to the detuning fluctuation $ \partial J/\partial \epsilon_0 \propto \theta^2$ is a second order effect.}

%in the basis ($\ket{T_0}$, $\ket{S^\prime}$),
The eigenstates of $H^\prime$ without noise can be obtained and denoted as $\ket{\uparrow\downarrow^{\prime\prime}}$ and $\ket{\downarrow\uparrow^{\prime\prime}}$ (see Figure \ref{Fig_scheme} (b)). In the basis of $\ket{\uparrow\downarrow^{\prime\prime}}$ and $\ket{\downarrow\uparrow^{\prime\prime}}$, the effective Hamiltonian including noise is
\begin{eqnarray}
H^{\prime\prime}&=&\left[\begin{array}{ccccc}
\frac{J}{2} + \frac{\Omega_J}{2} +\hat{n}_{\uparrow\downarrow^{\prime\prime},\uparrow\downarrow^{\prime\prime}} & \hat{n}_{\uparrow\downarrow^{\prime\prime},\downarrow\uparrow^{\prime\prime}} \\ %V_T and V_S is set to zero now
\hat{n}_{\downarrow\uparrow^{\prime\prime},\uparrow\downarrow^{\prime\prime}} & \frac{J}{2} - \frac{\Omega_J}{2} +\hat{n}_{\downarrow\uparrow^{\prime\prime},\downarrow\uparrow^{\prime\prime}} \\
\end{array}\right],
\end{eqnarray}
where $\Omega_J=\sqrt{J^2+\delta E_Z^2}$ is the energy splitting of $\ket{\uparrow\downarrow^{\prime\prime}}$ and $\ket{\downarrow\uparrow^{\prime\prime}}$,
%The angle $\theta_2$ of rotation that diagonalizes $H_0$ in the subspace $\ket{2}$ and $\ket{3^\prime}$ satisfies $\cos\theta_2=-J/\Omega_J$, and $\sin\theta_2=\delta E_Z/\Omega_J$.
$\hat{n}_{\uparrow\downarrow^{\prime\prime},\uparrow\downarrow^{\prime\prime}}= (1+J/\Omega_J)\hat{n}_{}^\prime/2$, $\hat{n}_{\uparrow\downarrow^{\prime\prime},\downarrow\uparrow^{\prime\prime}}=\hat{n}_{\downarrow\uparrow^{\prime\prime},\uparrow\downarrow^{\prime\prime}}=(\delta E_Z/\Omega_J)\hat{n}_{}^\prime /2$, $\hat{n}_{\downarrow\uparrow^{\prime\prime},\downarrow\uparrow^{\prime\prime}}= (1-J/\Omega_J)\hat{n}_{}^\prime/2$.

%Note: $\ket{\uparrow\downarrow}$ and $\ket{\downarrow\uparrow}$ are not product states!! they are entangled!

When $J \ll \delta E_Z $, which is satisfied in a two-qubit gate experiment \cite{veldhorst2015}, states $\ket{\uparrow\downarrow^{\prime\prime}}$ and $\ket{\downarrow\uparrow^{\prime\prime}}$ are approximately spin product states, where $\ket{\uparrow\downarrow^{\prime\prime}} \approx (\ket{(1,1)T_0} + \ket{(1,1)S^\prime})/\sqrt{2}$, $\ket{\downarrow\uparrow^{\prime\prime}} \approx (\ket{(1,1)T_0} - \ket{(1,1)S^\prime})/\sqrt{2}$, and the effective noise on $\ket{\uparrow\downarrow^{\prime\prime}}$ and $\ket{\downarrow\uparrow^{\prime\prime}}$ are $\hat{n}_{\uparrow\downarrow^{\prime\prime},\uparrow\downarrow^{\prime\prime}} \approx \hat{n}_{\downarrow\uparrow^{\prime\prime},\downarrow\uparrow^{\prime\prime}} \approx \hat{n}_{}^\prime/2$.
In this limit, the control-qubit and target-qubit are well defined. The control-qubit state effectively selects the subspace of the system (See Figure \ref{Fig_scheme} (b)).
If control-qubit is spin-down (spin-up), then, the system is in the subspace of $\ket{\downarrow\uparrow^{\prime\prime}}$ and $\ket{\downarrow\downarrow}$ ($\ket{\uparrow\uparrow}$ and $\ket{\uparrow\downarrow^{\prime\prime}}$).
%, which are coupled in experiments due to an applied AC magnetic field (not shown in our theory).
%Therefore, depending on the state of control-qubit, we are effectively selecting different TLSs with different level splittings.
%The exchange energy determines the difference of %the level spacings, i.e.
%Z-rotation speeds of target-qubit for different control-qubit states, and the Ramsey fringe can be utilized to show the state dependent CZ rotations and achieve the two-qubit logic gate. \cite{meunier_efficient_2011, veldhorst2015}
Thus, for a given state of the control-qubit, we can reduce the decoherence of a two-qubit system to the decoherence of a two-level system. We focus on pure spin dephasing, which is generally much faster than spin relaxation. %(see Appendix \ref{sec:T1})

\textbf{Decoherence formula---}
In this subsection, we develop the expression that determines the effect of noise.
For two states $\ket{\alpha}$ and $\ket{\beta}$ of interest,
%spin dephasing is determined by the spectral density
the system dephases as $\exp[{-\phi(\tau)}]$ \cite{duan_reducing_1998, taylor_dephasing_2006},
\beqa
\phi(\tau)=& \int_{\omega_0}^{\infty} d\omega J_{\alpha\beta}^{(zz)}(\omega) [2\sin(\omega \tau/2)/\omega]^2, \label{phit}\\
J_{\alpha\beta}^{(zz)}(\omega)=& \frac{2}{\hbar^2}\int_{-\infty}^{\infty}\langle \hat{h}_{\alpha\beta}^{(z)}(0)\hat{h}_{\alpha\beta}^{(z)}(\tau) \rangle \cos(\omega\tau)d\tau,
\eeqa
where %the correlation function is $\langle \hat{h}_{Z}^{(\alpha\beta)} (0)\hat{h}_{Z}^{(\alpha\beta)} (\tau) \rangle= \langle [\hat{n}^{\prime\prime}_{\alpha\alpha}(0)-\hat{n}^{\prime\prime}_{\beta\beta}(0)][\hat{n}^{\prime\prime}_{\alpha\alpha}(\tau)-\hat{n}^{\prime\prime}_{\beta\beta}(\tau) ]\rangle $
$\hat{h}_{\alpha\beta}^{(z)}=(\hat{n}_{\alpha,\alpha} -\hat{n}_{\beta,\beta})/2$ is the relative noise of the two states $\ket{\alpha}$ and $\ket{\beta}$ of interest, $J_{\alpha\beta}^{(zz)}(\omega)$ is the spectral density for the noise, and the cutoff frequency $\omega_0$ represents the inverse of the measurement time of coherence dynamics.

%the longitudinal noise
%The dephasing is determined by the longitudinal noise $\hat{h}_{\alpha\beta}^{(z)}$, which depends on the relative difference of noise on states $\ket{\alpha}$ and $\ket{\beta}$.

%$Since dephasing of TLS is determined only by the relative noise of the corresponding TLS, the decoherence of two-qubit system can be different from S-T$_0$ qubit.
Here, we emphasize the difference between a two-qubit gate system and a $S-T_0$ qubit before our detailed discussion of decoherence in the system.
A two-qubit gate system with two electrons in a DQD shares many similarities with a $S-T_0$ qubit in a DQD; however, there is an important difference.
The different is due to the fact that spin dephasing depends on the relative noise of two states rather than the noise of each individual states.
%(See Method)
%We denote $\hat{h}_{\alpha\beta}^{(z)}$ as the relative noise of the two states $\ket{\alpha}$ and $\ket{\beta}$ of interest.
For a $S-T_0$ qubit,  the qubit is encoded in states $\ket{\uparrow\downarrow^{\prime\prime}}$ and $\ket{\downarrow\uparrow^{\prime\prime}}$, the effective noise is $\hat{h}_{\uparrow\downarrow^{\prime\prime},\downarrow\uparrow^{\prime\prime}}^{(z)}=(\hat{n}_{\uparrow\downarrow^{\prime\prime},\uparrow\downarrow^{\prime\prime}} -\hat{n}_{\downarrow\uparrow^{\prime\prime},\downarrow\uparrow^{\prime\prime}})/2=(J/\Omega_J)\hat{n}_{}^\prime/2$. Increasing $\delta E_Z$, which reduces the ratio $J/\Omega_J$, can reduce the effective noise $\hat{h}_{\uparrow\downarrow^{\prime\prime},\downarrow\uparrow^{\prime\prime}}^{(z)}$ and suppresses the decoherence of $S-T_0$ qubit, as shown in a recent experiment \cite{nichol_high_fidelity_2017}. However, in a two-qubit gate system, for a given state of control-qubit, only one of $\ket{\uparrow\downarrow^{\prime\prime}}$ and $\ket{\downarrow\uparrow^{\prime\prime}}$ is involved. The relative noise will be either $\hat{n}_{\uparrow\downarrow^{\prime\prime},\uparrow\downarrow^{\prime\prime}}$ or $\hat{n}_{\downarrow\uparrow^{\prime\prime},\downarrow\uparrow^{\prime\prime}}$, which is not suppressed with increasing $\delta E_Z$. \textit{Therefore, in contrast to a $S-T_0$ qubit, spin decoherence in a two-qubit logic gate is not suppressed by increasing $\delta E_Z$.}

\textbf{Spectral densities---}
In order to study spin dephasing in the system, we need the corresponding spectra density. In a DQD, charge noise can induce detuning noise that arises from the non-identical noise on the two QDs, and tunneling noise that arise from fluctuations in barrier height.
When the control-qubit is initialized to be spin-down, the relevant two states are $\ket{\downarrow\uparrow^{\prime\prime}}$ and $\ket{\downarrow\downarrow}$. The effective noise is $\hat{h}_{\uparrow\downarrow^{\prime\prime},\downarrow\downarrow}^{(z)}=\hat{n}_{}^\prime/4$. (Dephasing for the spin-up control-qubit is {the same}, since $\hat{n}_{\uparrow\downarrow^{\prime\prime},\uparrow\downarrow^{\prime\prime}} \approx \hat{n}_{\downarrow\uparrow^{\prime\prime},\downarrow\uparrow^{\prime\prime}}$ when $J \ll \delta E_Z$.)
%, or
%\beq
%\hat{h}_{3^{\prime\prime}4}^{(z)}= \frac{1}{4}[\sqrt{2}{{\theta}}\hat{n}_t  - ({{\theta}}^2/4)\hat{n}_\epsilon ].
%\eeq
Since the charge noise is believed to be from noise-producing defects, homogeneously distributed in the plane of the device, fluctuation of the tunnel barrier height due to charge noise is of the same order as detuning fluctuations.
For non-correlated noises, the effective noise spectral density is given by (Supplementary information S2)
%(Supplementary information \ref{sec:Jzz34})
\beq
J_{\uparrow\downarrow^{\prime\prime},\downarrow\downarrow}^{(zz)}(\omega) =A_{eff}/\omega, \label{Jzz34}
%J_{3^{\prime\prime}4}^{(zz)}(\omega) =\frac{1}{8\hbar^2}\left[2{\theta}^2\left(\frac{\partial t_0}{\partial E_b}\right)^2 + \frac{\theta^4}{16}\right] S_{1/f}(\omega), \label{Jzz34}
\eeq
where $A_{eff}=\frac{A}{8\hbar^2}\left[2{\theta}^2\left({\partial t_0}/{\partial E_b}\right)^2 + \frac{\theta^4}{16}\right]$. The first term accounts for tunneling noise; the second for detuning noise.
$\partial t_0/\partial E_b$ converts the barrier fluctuation to fluctuations of the tunneling rate. In the WKB approximation,
\beq
\partial t_0/\partial E_b \approx t_0/(2\Delta_b), \label{dtdE}
\eeq
where $\Delta_b\equiv \sqrt{(E_b-E_0)\hbar^2/(2m^*l_b^2)}$, $E_0$ is the orbital energy of a single QD, $E_b$ and $l_b$ are the effective barrier height and width.
%Therefore, the effective noise spectral density is
%$J_{ZZ,3^{\prime\prime}4}(\omega)
%%\approx\frac{1}{\hbar^2}\left[{{\theta}}^2 S_{35}(\omega) +\frac{{{\theta}}^4}{16} S_{55}(\omega)\right],
%%{1}/{\hbar^2}\left[{{\theta}}^2 S_{35}(\omega) +{{{\theta}}^4} S_{55}(\omega)/{16}\right]
%=A_{eff}/\omega, \label{JZZ}
%$ where
%$A_{eff}= \frac{1}{8\hbar^2}[2{\theta}^2(\partial t_0/\partial E_b)^2 + {\theta}^4/16]A.$
With knowledge of $J_{\uparrow\downarrow^{\prime\prime},\downarrow\downarrow}^{(zz)}(\omega)$, the spin dephasing dynamics in $\exp[-\phi(\tau)]$ can be calculated from Eq. (\ref{phit}).

%The results of partially correlated and fully correlated noises are studied in the supplementary material.

%$A_{eff}= \frac{1}{8\hbar^2}[2{{\theta}}^2 t_0^2/(2\Delta_b)^2 + {{\theta}}^4/16]A$.
%$S_{35}(\omega)$ and  $S_{55}(\omega)$ are the spectral density of tunneling noise $\hat{n}_{t}$ and detuning noise $\hat{n}_{\epsilon}$, respectively, the noise correlation between $\hat{n}_{t}$ and $\hat{n}_{\epsilon}$ has been neglected for simplicity, and

Eq. (\ref{Jzz34}) and (\ref{dtdE}) indicate that the relative strength of $\sqrt{2}\partial t_0/\partial E_b \approx \sqrt{2}t_0/(2\Delta_b)$ and ${{\theta}}/4=\sqrt{2}t_0/(2(U-\epsilon_0))$ determines whether tunneling noise or detuning noise dominates. \textit{If $\Delta_b>U-\epsilon_0$, detuning noise dominates; if $\Delta_b<U-\epsilon_0$ tunneling noise dominates.}

\textbf{Spin decoherence results---}
%Before we show numerical results, we discuss the analytic dependence of spin dephasing on detuning based on Eq. \ref{Jzz34}.
The spin dephasing has $\exp[{-{A_{eff}}\tau^2\ln (1/(\omega_0 \tau))}]$ dependence (Supplementary information S3),
%(Supplementary information \ref{sec:analyticExpression}),
approximately $\exp[{-(\tau/T_\varphi)^2}]$ dependence, where $T_\varphi$ is the spin dephasing time.
We have evaluated Eq. (\ref{phit}) and fitted $\exp[{-\phi(\tau)}]$ to $\exp[{-(\tau/T_\varphi)^\beta}]$, and found that $\beta\lesssim 2$ (Supplementary information S4).
%(Supplementary information \ref{sec:dephasingDynamics}).
Therefore, the spin dephasing shows $1/T_{\varphi} \propto A_{eff}^{1/\beta}\approx \sqrt{A_{eff}}$ scaling, i.e. \textit{the dephasing rate $1/T_{\varphi}$ has approximately ${1}/{({U}-\epsilon_0)^2}$ dependence for detuning noise and ${1}/{({U}-\epsilon_0)}$ dependence for tunneling noise.}

\begin{figure}[t]
\centering
\includegraphics[scale=0.42]{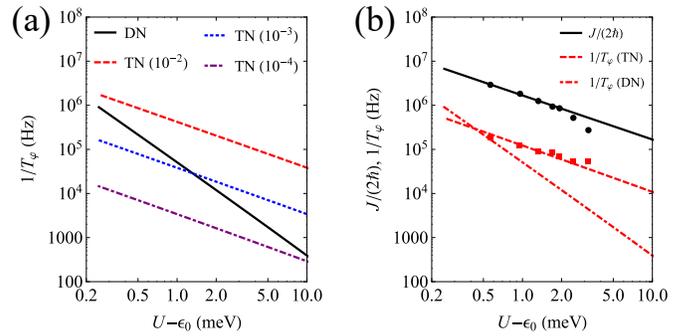}
\caption{Spin dephasing $1/T_\varphi$ as a function of detuning. (a):
$1/T_\varphi$ as a function of detuning $\epsilon_0$ due to tunneling noise (TN) only or detuning noise (DN) only. For tunneling noise, we choose representative values $\partial t_0/\partial E_b=10^{-2}$ (Dashed), $10^{-3}$ (Dotted), and $10^{-4}$ (Dashdotted).
(b): $J/(2\hbar)$ and $1/T_\varphi$ as a function of detuning $\epsilon_0$, where $A=(2 \mu eV)^2$ and $\partial t_0/\partial E_b=3.2 \times10^{-3}$.  {The dots are the experimental data for $J/(2\hbar)$ (circle) and $1/T_\varphi$ (square) \cite{veldhorst2015}.} }\label{fig:JTphi_detuning}
%The lines are the calculated results of $J/(2\hbar)$ (solid line), $1/T_\varphi$ due to tunneling noise only (dashed line) and $1/T_\varphi$ due to detuning noise only (dashdotted line)
%extracted from Ref.
%\caption{Pure dephasing rate $1/T_\varphi$ as a function of detuning $\epsilon_0$  due to detuning $1/f$ charge noise ($A=(1 \mu eV)^2$ and $t_0/(2\Delta_b)=10^{-3}$, and due to tunneling $1/f$ charge noise ($A=(1 \mu eV)^2$ and $\Delta_b=10^{-3}$) }\label{fig:Ncz_detuning}
\end{figure}

{
%For all $\epsilon_0$, we calculate the decoherence time $T_\varphi$ to be the time when the qubit phase coherence decays to $1/e$.
%, then, the decoherence rate can be obtained for various detunings of the DQD.
%the decoherence time $T_\varphi$ is about 3.6 $\mu s$ for the case of Figure \ref{fig:Gam22T_JN_a20}. Similarly,
%Figure \ref{fig:Gam33eps_tunnelingNoise} shows pure dephasing rate $1/T_\varphi$ of two-qubit gate system as a function of detuning $\epsilon_0$ of DQD due to tunneling $1/f$ charge noise while the control-qubit is spin down for different $\Delta_b$. The magnitude of $1/f$ charge noise $A=(1 \mu eV)^2$. The results shows the range of dephasing rate for several possible $t_0/\Delta_b$'s.
%Figure \ref{fig:Gam22eps_a20}
Figure \ref{fig:JTphi_detuning} (a) shows the detuning dependence of the dephasing rate of the target-qubit including only detuning noise and only tunneling noise (the control-qubit is spin-down). The dephasing due to detuning and tunneling noise show different detuning dependence, which enables the identification of different noise sources. For tunneling noise, results have been shown for different values of $\partial t_0/\partial E_b$, which modifies the crossover between tunneling noise and detuning noise. The dephasing due to tunneling noise can dominate over detuning noise in a wide range of detuning (note that $U=25$ meV).
%For even larger detuning, the dephasing could be dominated by detuning fluctuation, however, it happens only in a very small range of $\epsilon_0$ when $\epsilon_0$ is close to ${U}$.
}

{
Figure \ref{fig:JTphi_detuning} (b) shows a log-log plot of $J/(2\hbar)$  and dephasing rate $1/T_\varphi$ for only tunneling noise and only detuning noise. The experimental data shown as dots is extracted from Ref. \cite{veldhorst2015}. The calculated $1/T_\varphi$ due to detuning noise shows approximately ${1}/{({U}-\epsilon_0)^2}$ dependence, which is different from the experimental data, while $1/T_\varphi$ due to tunneling noise shows approximately ${1}/{({U}-\epsilon_0)}$ dependence. To match the experimental spin dephasing, we find $\partial t_0/\partial E_b=(3.2 \pm 0.2) \times10^{-3}$, which for WKB estimate $\partial t_0/\partial E_b \approx t_0/(2\Delta_b)\approx 4\times 10^{-4}$. We attribute the discrepancy to the simplicity of WKB method, the simplicity of the model barrier used, and the exponential dependence of tunneling on the parameters. (Note that the value of $\partial t_0/\partial E_b$ also depends on the choice of the amplitude $A$ of charge noise.) \textit{$J/(2\hbar)$ and $1/T_\varphi$ are almost parallel, indicating that they show the same $1/(U-\epsilon_0)$ dependence, and that the dephasing is dominated by the tunneling noise.} This parallel dependence doesn't change with variation of $\alpha_{cz}$ or $V_{cz0}$.

%, $A$, or $\partial t_0/\partial E_b$
}

{
%Detuning noise is usually assumed to be dominant. However, tunneling noise can be dominant.
The dominance of tunneling noise is counter to what is usually assumed.
To understand the qualitative behavior, we consider the WKB approximation Eq. (\ref{dtdE}).
Tunneling noise is dominant because $\Delta_b$ is small (lower tunnel barrier $E_b$ and bigger distance between QDs) compared to $U-\epsilon_0$ ($U$ is big in small dots). This tends to be satisfied in small silicon QDs using accumulation mode.
}

%\deleted{Eq. (\ref{JZZ}) shows that,
%if $\Delta_b \gg U-\epsilon_0$, detuning noise can dominate over tunneling noise; However,
%if $\Delta_b \ll U-\epsilon_0$, tunneling noise can be dominant. In the two-qubit {logic gate experiment}, the criteria is satisfied in most of the detuning regime.}

%When tunneling noise is dominant, the number of two-qubit operation $N_{CZ}=JT_{2}^*/(2\hbar)$ of CZ operations can increase as $\epsilon_0$ moves towards $U$ (more asymmetry) if a finite single qubit decoherence is considered in the total decoherence time $T_2^*$ (see Appendix \ref{sec:Ncz} for details). The weaker dependence of $1/T_{\varphi}$ with $1/({U}-\epsilon_0)$ compared to the case of detuning noise indicates that, by reducing $(U-\epsilon_0)$ half the magnitude, the two-qubit operation speed $J$ becomes twice faster, while the total decoherence rate $1/T_2^*$ is increased less if there is finite single qubit decoherence. There is always single spin decoherence due to mechanisms such as spin-orbit interaction. Thus, when tunneling noise is dominant, $N_{CZ}$ increases as $\epsilon_0$ approaches ${U}$, which is consistent with the experiment in Ref. \cite{veldhorst2015}.

The detuning dependence of the number of two-qubit operations can be different for detuning and tunneling noise.
When detuning noise is dominant, the dephasing rate $1/T_{2}^* \approx 1/T_\varphi \propto 1/(U-\epsilon_0)^2$, which increases faster than the exchange interaction ($J\propto 1/(U-\epsilon_0))$ as $\epsilon_0$ moves towards $U$ (more asymmetry). \textit{Thus,
when detuning noise is dominant, the number of CZ operations $N_{CZ}=JT_{2}^*/(2\hbar)$ reduces as the DQD becomes more asymmetric, as suggested in the recent experiments of a $S-T_0$ qubit in a GaAs DQD \cite{bertrand_quantum_2015, reed_reduced_2016, martins_noise_2016}.} However, when tunneling noise is dominant, $J$ and $1/T_\varphi$ have the same $1/(U-\epsilon_0)$ dependence, and the number of two-qubit operation $N_{CZ}$ is not reduced as DQD becomes more asymmetric. Furthermore, if there is finite spin dephasing $1/T_\varphi^{(1)}$ due to the single qubit mechanisms, then, the total dephasing rate $1/T_2^*\approx 1/T_\varphi + 1/T_\varphi^{(1)}$ has a slower scaling compared to $1/(U-\epsilon_0)$ (Supplementary information S5). \textit{Consequently, when tunneling noise is dominant, $N_{CZ}$ can increase as DQD becomes more asymmetric (i.e. $\epsilon_0$ approaches ${U}$), %(Supplementary information \ref{sec:Ncz}),
which is consistent with the experiment in Ref. \cite{veldhorst2015}.} This defines the regime of optimal operation when tunneling noise is dominant.

%The detuning dependence of the number of two-qubit operations can be different for detuning and tunneling noise.
%Suppose the total dephasing rate including single-qubit decoherence mechanisms is $1/T_2^*$, then, the number of two-qubit controlled-Z (CZ) gate (CPHASE gate when the phase is $\pi$) is $N_{CZ}=JT_{2}^*/(2\hbar)$.
%When detuning noise is dominant, the dephasing rate $1/T_{2}^* \approx 1/T_\varphi \propto 1/(U-\epsilon_0)^2$, which increases faster than the exchange interaction ($J\propto 1/(U-\epsilon_0))$ as $\epsilon_0$ moves towards $U$ (more asymmetry). \textit{Thus,
%when detuning noise is dominant, the number of CZ operations $N_{CZ}$ reduces as the DQD becomes more asymmetric, as suggested in the recent experiments of a $S-T_0$ qubit in a GaAs DQD \cite{bertrand_quantum_2015, reed_reduced_2016, martins_noise_2016}.} However, when tunneling noise is dominant, $J$ and $1/T_\varphi$ have the same $1/(U-\epsilon_0)$ dependence, and the number of two-qubit operation $N_{CZ}$ is not reduced as DQD becomes more asymmetric. Furthermore, if there is finite spin dephasing $1/T_\varphi^{(1)}$ due to the single qubit mechanisms, then, the total dephasing rate $1/T_2^*\approx 1/T_\varphi + 1/T_\varphi^{(1)}$ has a slower scaling compared to $1/(U-\epsilon_0)$ (Supplementary information). \textit{Consequently, when tunneling noise is dominant, $N_{CZ}$ can increase as DQD becomes more asymmetric (i.e. $\epsilon_0$ approaches ${U}$),
%which is consistent with the experiment in Ref. \cite{veldhorst2015}.}
%%(Supplementary information \ref{sec:Ncz}),

\section{Discussion}

Note that the value of $\partial t_0/\partial E_b$ we obtained for experiment depends on the choice of the amplitude $A$ of charge noise. If $A$ is chosen to be (1 $\mu eV$)$^2$ rather than (2 $\mu eV$)$^2$, then, we have $\partial t_0/\partial E_b=(6.4 \pm 0.4) \times10^{-3}$ to reproduce the results. Suppose that the amplitude of charge noise has been measured, then, the value of $\partial t_0/\partial E_b$ can be determined. Therefore, our theory also provides a metrology tool to further investigate the tunneling noises in those systems and characterize the system parameters, which is essential for the improvement of qubit behavior.

To reduce the effect of the tunneling noise, one needs to reduce the parameter $\partial t_0/\partial E_b$ as suggested by our theory. From a simple WKB approximation, one may reduce the parameter $\partial t_0/\partial E_b$ by using higher barrier and longer distance between two dots. Due to the valley physics, more detailed investigation will be needed in order to reduce $\partial t_0/\partial E_b$.

We should emphasize that the theory we developed here is general and can be applied to silicon, to spin qubits in GaAs QDs, to spin qubits on donor atoms, and even to hole spin qubit as long as other degrees of freedom are well separated from the problem of interest.
%We should emphasize that the theory we developed here is generally and can be applied to silicon, to spin qubits in GaAs QDs, to spin qubits on donor atoms, and even to hole spin qubit as long as other degrees of freedom are well separated from the problem of interest.
%We emphasize that the theory we developed here is not specific to silicon DQD, it can be generally applied to spin qubits in GaAs QDs, spin qubits on donor atoms, and even hole spin qubit as long as other degrees of freedom are well separated from the problem of interest.
Thus, tunneling noise from charge noise can have significant effect in those systems. The tunneling noise also has a similar effect on logical spin qubits, such as the $S-T_0$ qubit, which will affect the optimal operation of logical spin qubits.

In the discussion so far, we have assumed that the tunneling and detuning noises are uncorrelated. We do not think that it is likely that the noise will be correlated because the two types of noise are from different events at different locations. In the supplementary information. We also derived expressions for spectral density when the noise is partially and fully correlated, and discussed the consequences if the noise is correlated (Supplementary information S2 and S6). We find that, for the regime of the experiment, the tunneling contribution dominates and it does not matter whether or not the tunneling and detuning noises are correlated.

%\section{Conclusion}
%\textit{Conclusion---}
In conclusion, spin decoherence due to detuning and tunneling noises from $1/f$ charge noise is studied in a two-qubit gate system. The amplitude of tunnel noise is smaller than the detuning noise in general. However, the contribution of the detuning noise to spin decoherence is second order in the charge admixture, while the contribution of tunneling noise is first order.
As a consequence, decoherence due to tunneling noise can dominate over detuning noise for spin qubits in a DQD.
The different orders of contribution lead to different detuning dependence of spin dephasing, which enables the identification of the noise source. Decoherence is dominated by the tunneling noise from charge noise rather than detuning noise in a recent experiment of a two-qubit logic gate.
We identified the condition when tunneling noise dominates.
Furthermore, we find that when detuning noise dominates, symmetric operation indeed helps improve the number of two-qubit operation as suggested in recent experiments; however, this is not the case when tunneling noise dominates.
The results highlight the importance of tunneling noise and its {consequences on the optimization of} spin qubit operation.
The theory developed also provides a metrology method to investigate the tunneling noises and characterize the system parameters, which is essential for the improvement of qubit behavior.

%As a consequence, decoherence due to tunneling noise tends to dominate over detuning noise for spin qubits in small QDs using accumulation mode.
%In contrast to the symmetric operation to suppress spin dephasing from detuning noise, when tunneling noise dominates, more asymmetry can increase the number of two-qubit operation, which suggests the importance of considering tunneling noise to design optimal operation of spin qubit.
%We find that dephasing due to tunneling noise could dominate over detuning noise in the recent two-qubit gate experiment.
%When tunneling noise dominates qubit decoherence, more asymmetric detuning could increase the number of CZ operations.
%in small QDs using accumulation mode
%We find that the detuning noise is dominant when $\Delta_b\gg U-\epsilon_0$; and tunneling noise is dominant when $\Delta_b\ll U-\epsilon_0$. The

%Our theory also provides a metrology tool to investigate the tunneling noises and characterize the system parameters,
%which is essential for the improvement of qubit behavior.

\section{Methods}

%\textbf{Effective Hamiltonian---}
To study spin decoherence in the system, we first obtain an effective Hamiltonian. For a weak tunneling with $t_0\ll {U}-\epsilon_0$, the four lowest eigenstates are in the (1,1) charge configuration. In this limit,
%we can first consider the charge admixture of $\ket{S}$ and $\ket{S_D}$ due to the tunneling term in Eq. (\ref{H_C}). Thus,
we first diagonalize $H_C$ without noise and eliminate the double occupation state $\ket{(2,0)S}$. Then, an effective two-qubit Hamiltonian including the effect from charge noise is obtained (Supplementary information S1). From the effective Hamiltonian and spectral density of $1/f$ charge noise, the spin dephasing dynamics is evaluated.
\textbf{Parameters---}
We use similar parameters as in the experiment on a two-qubit gate \cite{veldhorst2015}.
The applied magnetic field $B_0=1.4$ T so the average Zeeman energy $\overline{E}_{Z}=0.162$ meV. The Zeeman energy difference $\delta E_Z=0.17$ $\mu$eV (40 MHz) due to a different g-factor modulation at each dot. The tunneling amplitude $t_0=2.63$ $\mu$eV (or 900/$\sqrt{2}$ MHz) [an extra factor of $1/\sqrt{2}$ due to a difference in the expression of $J$]. The onsite Coulomb energy (estimated from the charge stability diagram) ${U}=25$ meV.
The condition $t_0\ll {U}-\epsilon_0$ is always satisfied when ${U}-\epsilon_0 > 50$ $\mu$eV.
We choose a cutoff frequency $\omega_0=1$ $s^{-1}$, and an amplitude $A=$ (2 $\mu$eV)$^2$ for $1/f$ charge noise. %The parameters for charge noise will be discussed in the following sections.
To extract experimental data from Figure 3 of Ref. \cite{veldhorst2015},
%\deleted{we use $\alpha_{Q1}^{G1}=0.2$ ($\alpha_{Q2}^{G1}=0.05$) for the level arm of gate G1 to the left (right) qubit, and {$V_{G1}=19.6$ mV for $V_{CZ}=0$}.}
we convert the voltage $V_{cz}$ to detuning, $\epsilon_0 = \alpha_{cz} (V_{cz} + V_{cz0})$. We choose $\alpha_{cz}=0.19$ and $V_{cz0}=110.4$ mV, so that the calculated $J$ matches the experimental points in the detuning regime when $\epsilon_0$ is close to $U$.

\section{Data availability}

The main data supporting the finding of this study
are available within the article and its Supplementary Information files.
Additional data can be provided upon request.
%Supplementary information accompanies the paper is available on the npj Quantum Information website.

\section{Acknowledgment}
%The authors thank M. Veldhorst, A. Dzurak, X. Hu, and J. M. Taylor for many useful discussions.
The authors thank M. Veldhorst (U. Delft), A. Dzurak (UNSW), X. Hu (U. Buffalo), and J. M. Taylor (NIST) for useful discussions.

P.H. acknowledges the support by the Science, Technology and Innovation Commission of Shenzhen Municipality (No. ZDSYS20170303165926217, No. JCYJ20170412152620376)) and Guangdong Innovative and Entrepreneurial Research Team Program (Grant No. 2016ZT06D348) after joining in SUSTech.

\section{Competing interests} The authors declare no financial or non-financial conflicts of interest.

\section{Author contributions}

P.H. developed the theory and performed the calculations. G.B. supervised the project.
All authors researched, collated, and wrote this paper.

\end{document}